\documentstyle[12pt,epsfig]{article}
\begin{document}
\def\thefootnote{\fnsymbol{footnote}}
\setcounter{footnote}{1}
\begin{titlepage}
\hbox to \hsize{
\hfill\vtop{\hbox{}
%\vspace{5.mm}
\hbox{hep-ph/9603220}
%\vspace{2.mm}
\hbox{UB-HET-96-01}
%\vspace{2.mm}
\hbox{February 1996} } }
\vspace{20mm}
\begin{center}
{\Large \sc Anomalous Couplings and Chiral Lagrangians: \\[2.mm]
An Update\footnote{To appear in the Proceedings of the
LC2000 Workshop, Annecy, Gran Sasso and Hamburg, February -- September
1995}\\[1.cm]}

M.~Baillargeon$^1$, U.~Baur$^2$, G.~B\'elanger$^3$, F.~Boudjema$^3$, \\
G.~Couture$^4$, M.~Gintner$^5$ and S.~Godfrey$^5$\\[1.cm]

{\it$^1$GTAE, Lisbon, Portugal}\\
{\it$^2$Physics Department, SUNY Buffalo, Buffalo, USA}\\
{\it$^3$ENSLAPP-Annecy, France}\\
{\it$^4$Physics Department, UQAM, Montreal, Canada}\\
{\it$^5$Physics Department, Carleton University, Ottawa, Canada}\\[3.cm]
{\bf
ABSTRACT \\[6.mm]}
\begin{minipage}{5.5in}
{ \baselineskip=16pt 
We present an update of the limits expected for anomalous gauge boson
couplings in the language of chiral Lagrangian operators at the LHC and
the Linear Collider. Both, the $e^+e^-$ and the $\gamma\gamma$ mode of
the Linear Collider are analyzed. With a 500~GeV $e^+e^-$ collider,
and an integrated luminosity of 50~--~80~fb$^{-1}$, one reaches the domain
of precision measurements.
}
\end{minipage}
\end{center} 
\end{titlepage}

\newpage

\begin{center}
{\Large \bf Anomalous Couplings and Chiral Lagrangians: \\[1.mm]
An Update}

\vspace*{0.3cm}

\vspace*{0.3cm}

M.~Baillargeon$^1$, U.~Baur$^2$, G.~B\'elanger$^3$, F.~Boudjema$^3$, \\
G.~Couture$^4$, M.~Gintner$^5$ and S.~Godfrey$^5$

\vspace*{0.3cm}

\begin{tabular}{ll}
{\it$^1$GTAE, Lisbon, Portugal}\hspace*{.3cm}& 
%\hspace*{.1cm}
{\it$^2$SUNY Buffalo, Buffalo, USA} 
\\ 
{\it$^3$ENSLAPP-Annecy, France}\hspace*{.3cm} &
{\it$^4$UQAM, Montreal, Canada}\\
{\it$^5$Carleton University, Ottawa, Canada}&
%\hspace*{.1cm}
\end{tabular}
\end{center} 
\newcommand{\beq}{\begin{equation}}
\newcommand{\eeq}{\end{equation}}

\newcommand{\beqn}{\begin{eqnarray}}
\newcommand{\eeqn}{\end{eqnarray}}
 
\newcommand{\ra}{\rightarrow}
 
\newcommand{\su}{$ SU(2) \times U(1)\,$}
 
\newcommand{\gag}{$\gamma \gamma$ }
\newcommand{\gam}{\gamma \gamma }

\newcommand{\np}{Nucl.\,Phys.\,}
\newcommand{\pl}{Phys.\,Lett.\,}
\newcommand{\pr}{Phys.\,Rev.\,}
\newcommand{\prl}{Phys.\,Rev.\,Lett.\,}
\newcommand{\prep}{Phys.\,Rep.\,}
\newcommand{\zp}{Z.\,Phys.\,}
\newcommand{\sovjnp}{{\em Sov.\ J.\ Nucl.\ Phys.\ }}
\newcommand{\nuclinst}{{\em Nucl.\ Instrum.\ Meth.\ }}
\newcommand{\annp}{{\em Ann.\ Phys.\ }}
\newcommand{\intjmp}{{\em Int.\ J.\ of Mod.\  Phys.\ }}
 
\newcommand{\eps}{\epsilon}
\newcommand{\mw}{M_{W}}
\newcommand{\mww}{M_{W}^{2}}
\newcommand{\mwmw}{M_{W}^{2}}
\newcommand{\mhmh}{M_{H}^2}
\newcommand{\mz}{M_{Z}}
\newcommand{\mzz}{M_{Z}^{2}}

\newcommand{\lra}{\leftrightarrow}
\newcommand{\tr}{{\rm Tr}}
 
\newcommand{\dkg}{\Delta \kappa_{\gamma}}
\newcommand{\dkz}{\Delta \kappa_{Z}}
\newcommand{\dkgt}{$\Delta \kappa_{\gamma}\;$}
\newcommand{\dkzt}{$\Delta \kappa_{Z}\;$}
\newcommand{\dz}{\delta_{Z}}
\newcommand{\dgz}{\Delta g^{1}_{Z}}
\newcommand{\dgzt}{$\Delta g^{1}_{Z}\;$}
\newcommand{\la}{\lambda}
\newcommand{\lag}{\lambda_{\gamma}}
\newcommand{\laz}{\lambda_{Z}}
\newcommand{\lnl}{L_{9L}}
\newcommand{\lnr}{L_{9R}}
\newcommand{\lt}{L_{10}}
\newcommand{\lu}{L_{1}}
\newcommand{\ld}{L_{2}}

\newcommand{\ememt}{$e^{-} e^{-}\;$}
\newcommand{\epemt}{$e^{+} e^{-}\;$}
\newcommand{\epem}{e^{+} e^{-}\;}
\newcommand{\epemww}{e^{+} e^{-} \ra W^+ W^- \;}
\newcommand{\eewwt}{$e^{+} e^{-} \ra W^+ W^- \;$}
\newcommand{\epemwwt}{$e^{+} e^{-} \ra W^+ W^- \;$}
\newcommand{\ppwg}{p p \ra W \gamma}
\newcommand{\ppwz}{pp \ra W Z}
\newcommand{\ppwgt}{$p p \ra W \gamma \;$}
\newcommand{\ppwzt}{$pp \ra W Z \;$}
\newcommand{\gamgamt}{$\gamma \gamma \;$}
\newcommand{\gamgam}{\gamma \gamma \;}
\newcommand{\egamt}{$e \gamma \;$}
\newcommand{\egam}{e \gamma \;}
\newcommand{\gamgamwwt}{$\gamma \gamma \ra W^+ W^- \;$}
\newcommand{\gamgamwwht}{$\gamma \gamma \ra W^+ W^- H \;$}
\newcommand{\gamgamwwh}{\gamma \gamma \ra W^+ W^- H \;}

\newcommand{\ptu}{p_{1\bot}}
\newcommand{\vecptu}{\vec{p}_{1\bot}}
\newcommand{\ptd}{p_{2\bot}}
\newcommand{\vecptd}{\vec{p}_{2\bot}}
\newcommand{\ie}{{\em i.e.}}
\newcommand{\cm}{{{\cal M}}}
\newcommand{\cl}{{{\cal L}}}
\newcommand{\cd}{{{\cal D}}}
\newcommand{\cv}{{{\cal V}}}
\def\slashc{c\kern -.400em {/}}
\def\slashL{L\kern -.450em {/}}
\def\slashcl{\cl\kern -.600em {/}}
\def\W{{\mbox{\boldmath $W$}}}  
\def\B{{\mbox{\boldmath $B$}}}         
\def\noi{\noindent}
\def\nn{\noindent}
\def\sm{${\cal{S}} {\cal{M}}\;$}
\def\nph{${\cal{N}} {\cal{P}}\;$}
\def\sb{$ {\cal{S}}  {\cal{B}}\;$}
\def\ssb{${\cal{S}} {\cal{S}}  {\cal{B}}\;$}
\def\cviol{${\cal{C}}\;$}
\def\pviol{${\cal{P}}\;$}
\def\cpviol{${\cal{C}} {\cal{P}}\;$}

\newcommand{\lgg}{\lambda_1\lambda_2}
\newcommand{\lww}{\lambda_3\lambda_4}
\newcommand{\ppin}{ P^+_{12}}
\newcommand{\pmin}{ P^-_{12}}
\newcommand{\ppout}{ P^+_{34}}
\newcommand{\pmout}{ P^-_{34}}
\newcommand{\sinsq}{\sin^2\theta}
\newcommand{\cossq}{\cos^2\theta}
\newcommand{\yt}{y_\theta}
\newcommand{\hppll}{++;00}
\newcommand{\hpmll}{+-;00}
\newcommand{\hpplt}{++;\lambda_30}
\newcommand{\hpmlt}{+-;\lambda_30}
\newcommand{\hpptt}{++;\lambda_3\lambda_4}
\newcommand{\hpmtt}{+-;\lambda_3\lambda_4} 
\newcommand{\dk}{\Delta\kappa}
\newcommand{\klam}{\Delta\kappa \lambda_\gamma }
\newcommand{\kac}{\Delta\kappa^2 }
\newcommand{\lac}{\lambda_\gamma^2 }
\def\gamgamtzz{$\gamma \gamma \ra ZZ \;$}
\def\gamgamtww{$\gamma \gamma \ra W^+ W^-\;$}
\def\gamgamtwwe{\gamma \gamma \ra W^+ W^-}
\def\ggwwt{$\gamma \gamma \ra W^+ W^-\;$}

%\vspace*{1cm}

%TEXT BEGINS HERE

\section{Introduction}
\vspace*{-.3cm}
In case a Higgs boson is not found by the time the linear collider is
running, one will have to rely on studying the dynamics of the
Goldstone Bosons, {\it i.e.} how the 
$W$ and $Z$ interact, in order to probe the mechanism of symmetry 
breaking. 
In this scenario one expects the weak interactions to enter 
a new regime where the weak bosons become strongly interacting at 
effective energies 
of the order of a few TeV. At this scale new resonances might 
appear. However,  
the new dynamics could already be felt at lower energies through more 
subtle and indirect 
effects on the properties of the weak bosons. These effects may be 
revealed by precision measurements with the most prominent effects
appearing in the self-couplings of the $W$ and $Z$ bosons. 
Detailed investigations of these self-couplings 
therefore would provide a window to the mechanism of symmetry breaking. 

Present precision data leave no doubt about the local gauge 
symmetry~\cite{SchildknechtGI,Langacker95} while the proximity of
the $\rho$ parameter to one can be considered as strong evidence for a 
residual global (custodial) $SU(2)$ symmetry. 
Whatever the new dynamics may be, it should respect these 
``low energy'' constraints. 
In turn, this makes it possible to easily  parameterize possible effects 
of new physics 
in the $W$ sector by the introduction of a restricted set of 
higher order operators. Such constructions  have been discussed at 
length~\cite{Hawai,BDV} and therefore, in this short note, we  
we will only recall the minimal set of operators that describe the
residual effect of the new dynamics in the absence of Higgs boson field:
\beqn
{{\cal L}}_{9R}=-i g' \frac{L_{9R}}{16 \pi^2} \tr (  \B^{\mu \nu}\cd_{\mu}
\Sigma^{\dagger} \cd_{\nu} \Sigma ) & &
{{\cal L}}_{9L}=-i g \frac{L_{9L}}{16 \pi^2} \tr ( \W^{\mu \nu}\cd_{\mu} 
\Sigma \cd_{\nu} \Sigma^{\dagger} )  \nonumber \\
\cl_{1}=\frac{L_1}{16 \pi^2} \left( \tr (D^\mu \Sigma^\dagger D_\mu 
\Sigma) 
\right)^2 \;\;\;\;\;\;\;\;\; & &
%$\;\;\;\;\;\;\;\;---------\;\;\;\;$&$
\cl_{2}=\frac{L_2}{16 \pi^2} \left( \tr (D^\mu \Sigma^\dagger D_\nu 
\Sigma)
\right)^2  
\eeqn
The first two operators contribute to the tri-linear couplings \dkgt
and \dkzt. Only $L_{9L}$ affects \dgzt.  
$L_{1,2}$ contributes solely to the quartic couplings (see for
instance~\cite{Hawai}). 
There is, in fact, yet another operator in conformity with the above 
symmetries: 
\beqn
{{\cal L}}_{10}&=&g g' \frac{L_{10}}{16 \pi^2} \tr ( \B^{\mu \nu}
\Sigma^{\dagger} \W^{\mu \nu}  \Sigma ) 
\eeqn
However, because it contributes to the two-point function, it is
strongly constrained through the $S$ parameter~\cite{Peskin} by 
LEP data: $ L_{10}=-\pi S$. Current limits~\cite{Langacker95} lead to 
\beqn
-0.7<L_{10}<2.4
\eeqn 
It will be very difficult to improve this limit in experiments at
future colliders. This poses a naturalness problem, since one would expect 
the other operators to be of the same order~\cite{Ruj}. If this were 
the case the improvement 
that might result from high energy colliders would, at best, be marginal. 
For instance, such a situation is encountered with a naive 
scaled-up-QCD technicolour. One way out is to associate the smallness of 
$L_{10}$ to a symmetry that forbids its appearance, 
in the same way that the custodial $SU(2)$ symmetry prevents large
deviations of the $\rho$ parameter from one. $L_{10}$ represents the 
breaking of the axial global $SU(2)$ symmetry~\cite{InamiL10}. 
Models that naturally incorporate the $L_{10}$ 
constraint include dynamical vector models that deviate from the usual 
scaled up versions of 
QCD by having (heavy) degenerate vectors and axial-vectors like the 
extended 
BESS model~\cite{BESS}. The latter implements an $(SU(2)_L \times 
SU(2)_R)^3$ symmetry. In the following we will follow a model
independent approach, assuming that the couplings of Eq.~(1) are
independent parameters. Taking $L_{10}\sim 0$ we investigate 
whether future machines could do as well as LEP1 in constraining the 
remaining operators. 

\section{Tri-linear Couplings: $L_{9L}, L_{9R}$}

In the past, extensive studies on the extraction on the anomalous 
tri-linear couplings have been performed.  
Our aim here is to update some of those results for the particular case 
of the chiral lagrangian approach (Eq.~(1)).  

\subsection{Strategy and Analysis at the NLC}
At the NLC, the best channels to look for $L_{9L,R}$ are  \epemwwt and 
\gamgamwwt. In the \epemt mode one could also use $\epem \ra \nu_e 
\bar\nu_e \gamma$~\cite{eenng} 
(which singles out the photonic part) or $\epem \ra \nu \bar\nu 
Z$~\cite{eennz} (which 
isolates the $WWZ$ part). Beside the fact that within the chiral 
Lagrangian approach both non-standard $WW\gamma$ and $WWZ$ couplings 
appear, the latter channels are found not to be competitive 
with $W$ pair production. One reason is that $W$ pair production has 
a much richer helicity structure that can directly access the 
longitudinal modes of the $W$. $WW\gamma$ and $WWZ$~\cite{nousee3v} 
production in \epemt collisions are quite interesting but, as far as 
tri-linear couplings are concerned, they can compete with the $WW$ 
channel only for TeV energies since they 
suffer from much lower rates~\cite{nousee3v}. However they are well suited  
to study possible quartic couplings. In this respect $WWZ$ production 
can probe the all important ${\cal L}_{1,2}$ operator.

In the \ememt mode one can investigate the tri-linear couplings 
through $e^-e^- \ra e^- W^-\nu$ \cite{CuypersenW}, which unfortunately 
exhibits the same shortcomings as  $\epem \ra \bar \nu_e \nu_e Z$. 
Cuypers has studied 
the potential of this reaction in probing the tri-linear couplings by 
taking into account 
the possibility of polarized beams, fits have only been done to the 
scattering 
angle of the  final electron. The limits do not compare well with those 
obtained from \epemwwt. 
\noi The $\gamma \gamma$ mode is ideal in probing the photonic 
couplings due to the very large cross section for $WW$ 
production~\cite{nousggvvref}. In the chiral approach this 
reaction will only constrain the combination $L_{9L}+L_{9R}$ but, as we 
shall see, in conjunction with the \epemt mode this is very helpful. 

$W$ pair production, both in \gamgamt and \epemt collisions, 
provides a large data sample and involves different helicity states. 
To fully exploit these reactions it is then important to reconstruct 
all the elements of the density
matrix for these reactions (both in the \gamgamt and \epemt mode). This 
can be achieved by analysing all 
the information provided by the complete set of the kinematical 
variables  related to the decay
products of the $W$'s, rather than restricting the analysis to the angular 
distribution at the level of the $W$. Since binning in the 5 variables
characterizing a $WW\to 4$~fermion event requires high statistics, a 
$\chi^2$ fit is not very efficient and a maximum likelihood technique is
used. Initial state polarization can also be easily implemented. The 
results presented here are based on using the 
semi-leptonic final states only. The impact of the 
the non-resonant diagrams (which could introduce a bias) is also 
quantified. 
The issue of luminosity and the improvement in the limits by going to 
higher energy will be discussed, and we shall compare our results with
those of Barklow~\cite{Barklow} which include initial state radiation
(ISR) effects.

\subsection{Analysis at the LHC}

Before discussing the limits on the parameters of the chiral Lagrangian 
that one 
hopes to achieve at the different modes of the linear colliders it is 
essential 
to compare with the situation at the LHC. For this we assume the high 
luminosity option with 100~fb$^{-1}$. The LHC limits are based on a very 
careful
study~\cite{Baur} that includes the very important effects of the QCD
NLO corrections as well as implementing the full spin correlations for the 
most interesting channel $pp \ra WZ$. $WW$ production 
with $W\ra jets$ production is fraught with a huge QCD background, while 
the leptonic mode is extremely difficult to reconstruct due to the two 
missing neutrinos. Even so, a thorough investigation (including NLO QCD 
corrections)  
for this channel has been done~\cite{Baurnloww}, which confirms the 
superiority of the $WZ$ channel. The NLO corrections for $WZ$ and $WW$ 
production at the LHC are huge, especially at large $W$ and $Z$ boson
transverse momenta where effects of the anomalous couplings are
expected to show up. 
In the inclusive cross section this is mainly due to, first, the 
importance 
of the subprocess $q_1 g\ra Z/W q_1$ (large gluon density at the LHC) 
followed by the ``splitting'' of the quark $q_1$ into $W/Z$. The 
probability for this splitting
increases with the $p_T$ of the quark (or $Z/W$): Prob$(q_1\ra q_2 W) \sim
\alpha_w/4\pi \ln^2(p_T^2/M_W^2)$. To reduce this effect one has to define 
an exclusive cross section that should be as close to the LO $WZ$ cross 
section as possible by cutting on the extra high $p_T$ quark, rejecting 
any jet with 
$p_T^{{\rm jet}}>50$~GeV, $|\eta_{{\rm jet}}|<3$. This defines a NLO 
$WZ/WW +``0$~jet'' cross section which is stable against variations 
in the choice of the $Q^2$ but which nonetheless can
be off by as much as $20\%$ from the prediction of the Born \sm result. 

\subsection{Comparison and Discussion}
\begin{figure*}[htb]
\caption{\label{newl9.fig}{\em Limits on ($L_{9L},~L_{9R}$) from
\epemt collisions including ISR and beam polarization effects with only 
the resonant diagrams.
The effect of keeping all resonant diagrams for the semi-leptonic final 
state 
is also shown. Limits from $\gamma \gamma \ra W^+ W^-$, and $WZ$ and $WW$
production at the LHC, are also shown for comparison. }}
%\cite{Barklow}
\begin{center}
\vspace{-1cm}
%\mbox{\epsfxsize=8cm\epsfysize=8cm\epsffile{barklow.eps}}
\hspace*{1.5cm}\mbox{\epsfxsize=13.5cm\epsfysize=15cm\epsffile{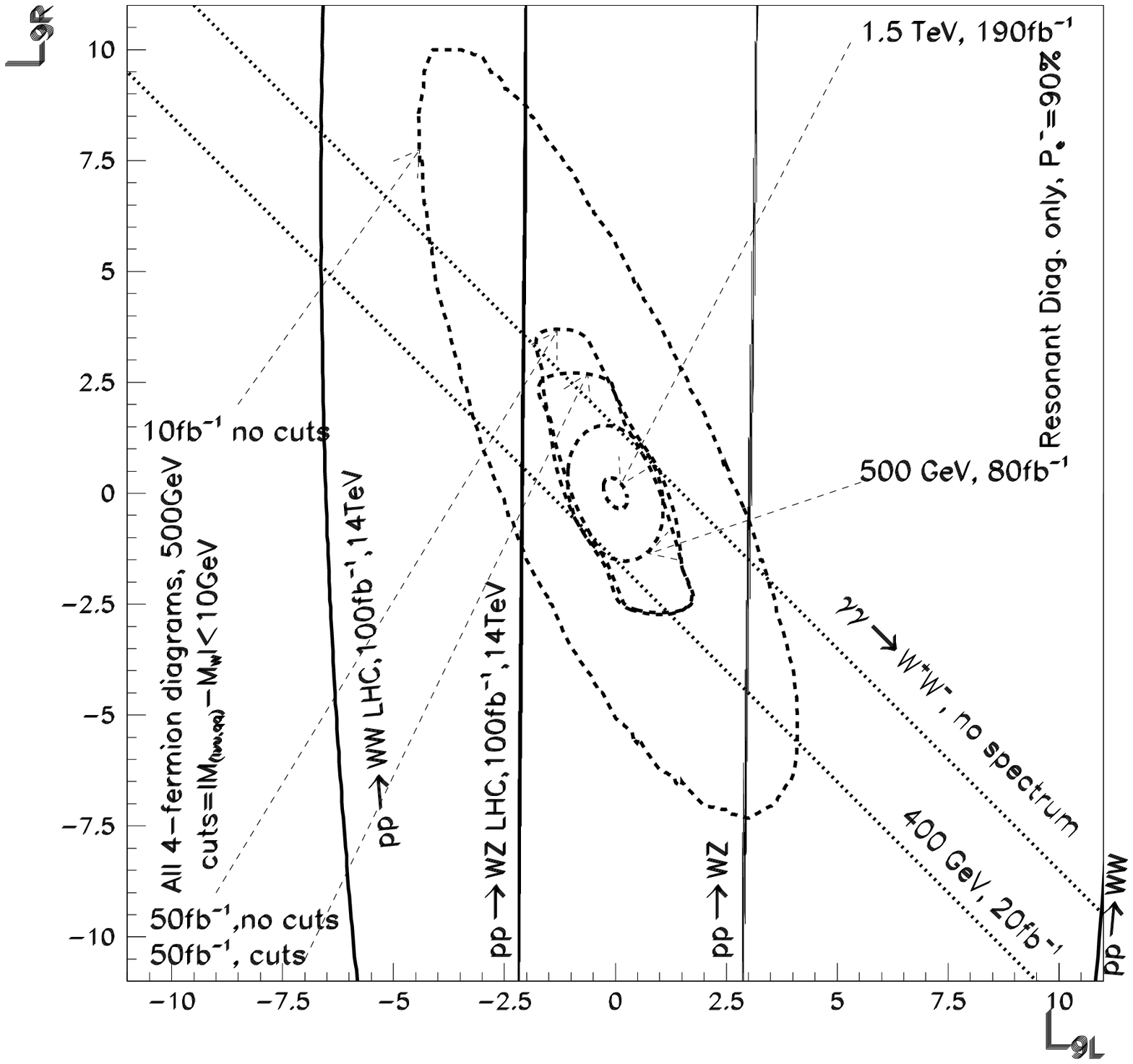}}
\vspace*{-1cm}
\end{center}
\end{figure*}
%\vspace*{0.5cm}
Fig.~\ref{newl9.fig} compares the limits one expects from the NLC 
and LHC while quantifying the various approximations that 
can affect the \epemwwt analysis. First, we address the issue of 
non-resonant contributions. Gintner, Godfrey and Couture~\cite{GGC} 
considered all 
diagrams that contribute to to the semi-leptonic $WW$ final state.
Requiring that the $jj$ and $\ell\nu$ invariant masses are both within
10~GeV of $M_W$, one essentially selects the doubly-resonant, $WW$ 
mediated, process. Taking into account all contributions 
including the background to $WW$ production only marginally 
degrades the limits. Changing the luminosity can to a very 
good approximation be accounted 
for by a scaling factor, $\sim \sqrt{{\cal L}}$ (compare the analyses 
performed with an integrated luminosity of 10~fb$^{-1}$ and 50~fb$^{-1}$). 
This result is confirmed also by the analysis conducted by 
Barklow~\cite{Barklow} which assumes higher luminosities but incorporates 
ISR effects as well as beam polarization. When polarization is assumed, 
the total luminosity shown on the plot is shared equally between a 
left-handed and a right-handed electron (assuming $90\%$ longitudinal 
electron polarization).  

The important conclusion drawn from Fig.~\ref{newl9.fig} is that 
through \epemwwt alone one indeed reaches the domain of 
precision measurements with 
an integrated luminosity of 50~--~80~fb$^{-1}$, matching the accuracy
of LEP1 for $L_{10}$. It is quite fascinating that 
we can achieve this level of precision already with $\sqrt{s}=500$~GeV. 
Moving to the TeV range, the limits improve by an additional order of 
magnitude (see Fig.~\ref{newl9.fig}).

\gamgamwwt is quite important. 
The results~\cite{nousggvvml} shown in Fig.~\ref{newl9.fig} 
consider a luminosity of 20~fb$^{-1}$ with a peaked fixed spectrum 
corresponding to a center of mass \gamgamt energy of 400~GeV ($80\%$ of
$\sqrt{s_{ee}}=500$~GeV). 
Convolution with a photon spectrum has not been done, since recent 
studies show that 
there is still much uncertainty concerning the form of the spectrum due 
to multiple rescattering effects~\cite{Schulte}. 
Clearly the \gamgamt mode helps to improve the limits extracted for
\epemt collisions at $\sqrt{s}=500$~GeV and integrated luminosities of 
50~fb$^{-1}$ or less. 

Fig.~\ref{newl9.fig} also compares the situation with the LHC. 
One observes that the limits obtained from $W^\pm Z$ production are
considerably better than those derived from the $WW$ channel. 
This is mostly due to the absence of serious background contributions in
the $WZ$ case. In the $WW$ case, $t\bar t$ production is the main
background which is difficult to suppress~\cite{Baurnloww}.
However, since $pp\to WZ$ effectively only constrains $L_{9L}$ (through 
$\Delta g_1^Z$), the LHC is not very sensitive to $L_{9R}$. 
As a result, with 50~fb$^{-1}$ and 500~GeV,  
the NLC constrains the two-parameter space much better than the LHC. 

Finally, we briefly comment on
the genuine quartic couplings, which are parameterized through $L_{1,
2}$. These are extremely important as they are the only couplings which
involve the longitudinal modes and hence are of crucial relevance when 
probing the Goldstone interaction. They are best probed through $V_L
V_L \ra V_L V_L$ scattering. However, for $\sqrt{s}=500$~GeV the 
$V_L$ luminosity inside an electron is unfortunately rather small, and 
one has to 
revert to $\epem \ra W^+W^-Z$ production, as suggested in~\cite{nousee3v}. 
This channel has been re-investigated by 
A.~Miyamoto~\cite{Miyamoto2} who conducted a detailed simulation 
including $b$-tagging to reduce 
the very large background from top pair production. With a luminosity of 
50~fb$^{-1}$ at 500~GeV, the limits are not very promising and do not pass 
the benchmark criterium $L_i<10$. It is found that $-95< L_1<71,\;\;-103< 
L_2<100$ 
(one parameter fits). These limits agree very well with the results of a 
previous analysis~\cite{nousee3v}. To seriously probe these special 
operators one needs energies in excess of 1~TeV. At 1~TeV the bounds 
improve to $L_{1,2} \sim 6$~\cite{Hawai}. However, it is difficult to 
beat the LHC here, where limits of  
${\cal O}(1)$ are possible~\cite{BDV} through $pp \ra W^+_L W^+_L $. 

In conclusion, it is clear that already with a 500~GeV \epemt collider
and an integrated luminosity of about 50~--~80~fb$^{-1}$ one can reach a 
precision on the parameters that probe \sb in the genuine tri-linear $WWV$
couplings which is similar to that which
can be achieved with LEP1 from oblique corrections to the $Z$ boson
parameters. The sensitivity of the NLC is further enhanced 
if \ggwwt can be studied.

Results from this channel would provide
invaluable information on the mechanisms of symmetry breaking, 
if no new particle is observed at the LHC or NLC (Light Higgs and 
SUSY). The NLC is unique in probing the vector models that 
contribute to 
$L_9$ (with $L_{1,2}\sim 0 $) and hence is complementary to the LHC. The 
latter is extremely efficient 
at constraining the ``scalar'' models. To probe deeper into the structure 
of symmetry breaking, a linear collider with  an energy range $\sqrt{s} 
\geq1.5$~TeV  would be most welcome.  
\vspace*{0.5cm}

\noindent{\large \bf Acknowledgements}

We would like to thank Tim Barklow for providing us with the results of his
analysis. 

%\vspace*{1.cm}

\end{document}